\documentclass[11pt, a4paper]{article}
\usepackage{epsfig}
\usepackage{graphicx}
\usepackage{amsmath}

\usepackage{doublespace}
\usepackage{amssymb}
\begin{document}

\title{THE FRACTAL DIMENSION OF AN OIL SPRAY}
\author{{\bf R. Castrej\'on Garc{\'\i}a$^1$, A. Sarmiento Gal\'an$^2$,} \and 
{\bf J. R. Castrej\'on Pita$^3$ and A. A. Castrej\'on Pita$^3$} \\
%EndAName
$^1${\small Instituto de Investigaciones El\'ectricas, Av. Reforma 113, 62490 
Cuernavaca, Morelos, M\'exico.} \\
$^2${\small Instituto de Matem\'aticas, UNAM, Av. Universidad s/n, 62200 
Chamilpa, Morelos, M\'exico.} \\
$^3${\small Centro de Investigaci\'on en Energ{\'\i}a, UNAM, Apdo. Postal 34, 
62580 Temixco, Morelos, M\'exico.}}
\maketitle

\begin{abstract}
We study the fractal dimension of the contour of the liquid-gas interface in 
a spray. Our images include both, the linking region and the break-up region 
and are obtained with a high-resolution shadowgraph technique; this means that 
the images can then be subject to an intensity filtering, equivalent to a 
threshold analysis, that enables the establishment of the fractal range. 
\end{abstract}

\vfill\eject

\section{Introduction}

The study of the geometric complexity in the break-up region of the liquid-gas 
interface in sprays represents a topic of special interest due to the fact 
that the physical process of atomization\footnote{We use here the common, 
albeit incorrect, sense for the word. According to its etymology, atom is an 
indivisible entity, and thus, to atomize really means to make indivisible, a 
meaning which is exactly the opposite to the intention when commonly used.} is 
intimately related to many commercial and technological 
applications~\cite{PRE}. The use of aerosol visualization analysis includes 
fields as diverse as the application of pesticides and insecticides to crops, 
hand-held sprays for painting or drying, fuel injection in combustion engines, 
liquid fuel spraying in the power generation industry, spray coating of 
pharmaceutical products, nuclear core cooling, and commercial filling 
processes, to mention just a few~\cite{Baker,Handbook}, and the studies are 
carried out with the efficient application of liquids, fuels or oils, in mind. 
In our study, the emphasis is set on the details of the relation between some 
properties of the fluid, like homogeneity or terminal speed, and the distance 
to the spray nozzle.

\noindent The use of shadowgraph images as a visualization tool is not new; 
their use in complexity studies however, has not been thoroughly exploited. 
Since our images are obtained with a high-resolution shadowgraph technique, 
they have proven specially pertinent for the implementation of the detailed 
analysis presented elsewhere for experiments in transport by 
fluids~\cite{Prassad}.

\section{Experimental Setup}

Among the different optical techniques that are used for the study of 
particles, liquids or gases in motion, the shadowgraphy stands out as an 
inexpensive and powerful technique~\cite{Jones}. Its main advantage is to 
highlight the relative refractive index of bodies, and thus to allow the 
visualization of objects that would be impossible to attain with the 
conventional photographic techniques. This is particularly useful when the 
objects to photograph are transparent instead of opaque, {\it i. e.}: air and 
water, or transparent fluids with different refractive indexes.

\noindent In the shadowgraph technique the object is illuminated from its back 
through an optical system designed specifically for such purpose. In this way, 
the light that strikes the photographic film is the one that passes through, 
without interacting with the object; the light that interacts with the object 
is absorbed or deviated aside due to a difference in its refractive index and 
does not influence the film. Figure 1 shows a schematic view of the 
shadowgraph system used for this work.

\noindent The atomization process occurs as a result of the interaction 
between the oil in its liquid state and the surrounding air, and involves 
several stages through which the oil becomes an aerosol. These stages can be 
clearly observed in any of the pictures. Initially, as soon as the oil is 
forced to leave the atomizer nozzle with a velocity that has both axial and 
radial components, the fluid is turned into thin laminar waves that are 
gradually converted by the aerodynamic forces into thin ligaments. Downstream, 
these ligaments are broken up into a cloud of small droplets that continue to 
move with a terminal average velocity of several meters per second.

\noindent Our photographs correspond to the spray of transparent oil produced 
by a pressure swirl atomizer: the nozzle orifice is $1$ mm in diameter and 
produces a hollow cone spray with an amplitude of $45^{\circ}$ (Fig. 2). The 
atomization pressure, obtained with a common gear pump, is $4 \times 10^5$ Pa. 
The fluid is a blend of industrial oils designed to match the characteristics 
of diesel fuel which produces an impressive and complex droplets formation 
spray~\cite{Bombrowski}.

\noindent The measurement of the terminal speed was carried out with a 
technique preceding the Particle Image Velocimeter (PIV) based on a double 
photographic exposure. The break-up region is optically magnified (three times 
in our case) and the ISO $400$ b\&w professional film is imprinted by a double 
flash from an Argon jet stabilized spark gap which lasts for $300$ ns and is 
shot again after $0.1$ ms; this creates a double image whose characteristic 
and identifiable features are separated by a certain distance that, together 
with the time lapse between flashes and the optical magnification, allow for 
the determination of the droplets speed. The mean speed thus measured at the 
break-up region, $30$ mm from the nozzle, has a value of $15.5$ m/s. Although 
no magnification exists in the photographs used for this work, the type of 
film employed allows for a photographic magnification of up to twenty times 
without any sharpness loss. Finally, in order to analyze their structure, the 
photographs were digitalized with a resolution of $150$ dpi in an 8-bit 
gray-scale comprising $254$ tonalities of gray.

\section{Measurements}

We have used the well-know box-counting method to calculate the fractal 
dimension of the spray contour: a grid of size $\varepsilon $ is superimposed 
on the black and white images and then, the boxes of this grid that intersect 
any part of the image are counted to provide the number $N$; the process is 
repeated reducing the size of the grid and thus, the discrete function 
$N(\varepsilon )$ is generated. The initial size is usually a one pixel 
grid~\cite{Castrejon,Harfa,Fractals} and the number of pixels of size 
$1/\varepsilon$ is increased until the grid size reaches a certain previously 
determined value (this value is usually given by the smallest possible pixel 
size). Finally, one is interested in the limit

\begin{equation}
D \ = \ \lim_{\varepsilon \to 0} \ \frac{\ln N(\varepsilon )}
{\ln (1/\varepsilon )}\nonumber
\end{equation}
which defines the fractal dimension $D.$ For the analysis of the contour of 
the oil aerosol, examples shown in Figs. 3 and 4, we varied the size of the 
grid from $1$ to $1,300$ pixels (the longitude of the image in pixels) and 25 
different images of the spray were taken under the same physical conditions; 
within the attainable accuracy, the corresponding results produce all the same 
value.

\noindent The box-counting method is designed for monochromatic 
images~\cite{Harfa,Yorke} and thus one has to be very careful when analyzing 
either color or varying intensity images. This is due to the fact that the 
features from the image whose intensity is below a certain value, called the 
threshold level, will be lost when transforming the image to a black and white 
one~\cite{Prassad}. In our case, this problem is represented by some parts of 
the image being less opaque than others due to light extinction; some thin 
filaments are less opaque than their thicker similes. Even though every 
precaution was taken to adjust the depth of field to the size of the spray, 
some small contribution to this intensity problem may arise from a lack of 
perfect focusing.mTo avoid complications coming from this difficulty, our 
analysis was carried out by filtering each image at every gray tonality from 
the $0^{th}$ to the $254^{th}$ level of the digitalized images, {\it i. e.}, 
the first filter allows only for the black parts to be taken into account, and 
then the intensity is diminished one digital step to incorporate gradually the 
less opaque features of every image. If this filtering process is not carried 
out, the simple transformation of the images to black and white ones produces 
the equivalent of filtering every image at half its maximum intensity value, 
Fig. 6. However, as shown in the following section, the filtering process just 
described reveals the rich fractal structure of the oil spray over a much 
wider range of intensity values or gray tonalities. 

\section{Results}

The fractal dimension of the contour of the oil spray when the images are only 
filtered at half their maximum intensity is $1.57 \pm 0.03$ with a linear 
correlation coefficient of $0.98$; the result is illustrated in Fig. 6. The 
whole spectrum of the fractal dimension of the contour however, is shown in 
Fig. 7, where the diameter of the open circles indicates the uncertainty at 
each intensity value and corresponds to the standard deviation of the data in 
all the $25$ analyzed photographs. To illustrate the fractal structure and the 
self-similarity of the contour at different intensities, the graph in Fig. 7 
includes the fractal dimension value obtained from the filtering process 
starting at the level in which the image becomes a single black rectangle up 
to the level of a single white one. The graph displays the fractal structure 
for the whole digital intensity range; a small star indicates the already 
mentioned fractal dimension at half-intensity filtering ($126{th}$ digital 
level). A linear fit with a correlation coefficient of $0.997$ is also shown, 
the slope and ordinate are: $(9.969 \pm 0.154) \times 10^{-4}$ and $1.436 \pm 
0.002$, respectively. After the $249^{th}$ digital level, the filtering leaves 
no tractable structure in the image; the gradual decay of the fractal 
dimension observed in Fig. 7, from the $249^{th}$ to the $254^{th}$ digital 
level of gray, is therefore due to a tenuous gray intensity present in the 
background of our photographs. 

\section{Conclusions}

The shadowgraph technique allows to obtain high-definition images of the whole 
oil spray which are ideal for the analysis of the dynamics of the aerosol and 
even of its smallest details since the image can be magnified several tenth 
times without noticeable sharpness losses; a higher density of dots is 
equivalent to a smaller pixel size which enables the analysis of smaller 
regions in the image. 

\noindent For complexity studies of similar systems through fractal dimension 
analysis~\cite{Yorke}, it may be possible to use b\&w photographs since the 
strong linear relation between intensity and threshold assures the existence 
of an interval (centered at the b\&w intensity value) where the fractal 
dimension will remain linear in relation to the intensity. However, the 
analysis carried out for the contour intensity clearly indicates that, in this 
and similar cases, it is not suitable to directly digitalize the images in a 
purely black and white format since then all the rich structural information 
contained in the gray tonalities is irremediably lost.

\vskip2cm

\footnoterule{\noindent This work has been partially supported by the IIE 
(M\'exico), and DGAPA-UNAM (project IN101100); AACP acknowledges financial 
support from DGEP-UNAM}

\vfill\eject

\vfill\eject

\null\vskip3cm

\begin{figure}[h]
\begin{center}
\epsfig{width=12cm,file=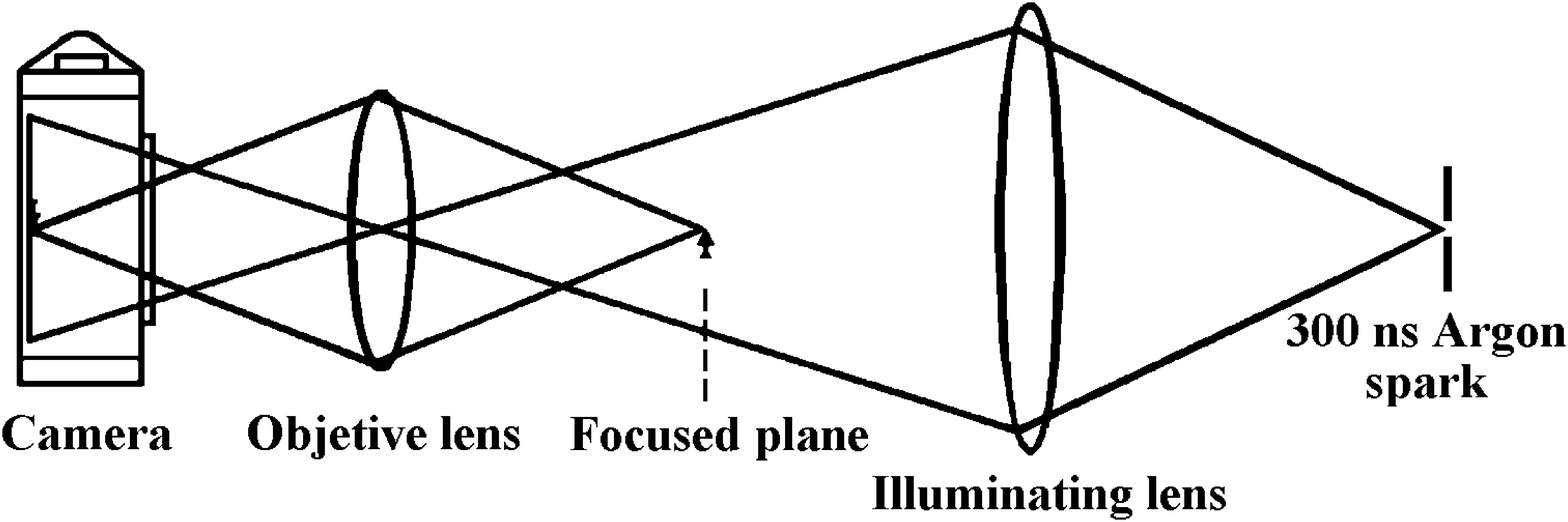}
\vskip-0.1cm
Fig. 1 ~Schematic view of the shadowgraph device used.
\end{center}
\vskip-0.4cm
\end{figure}

\vskip3cm

\begin{figure}[ht]
\begin{center}
\epsfig{width=5.cm,file=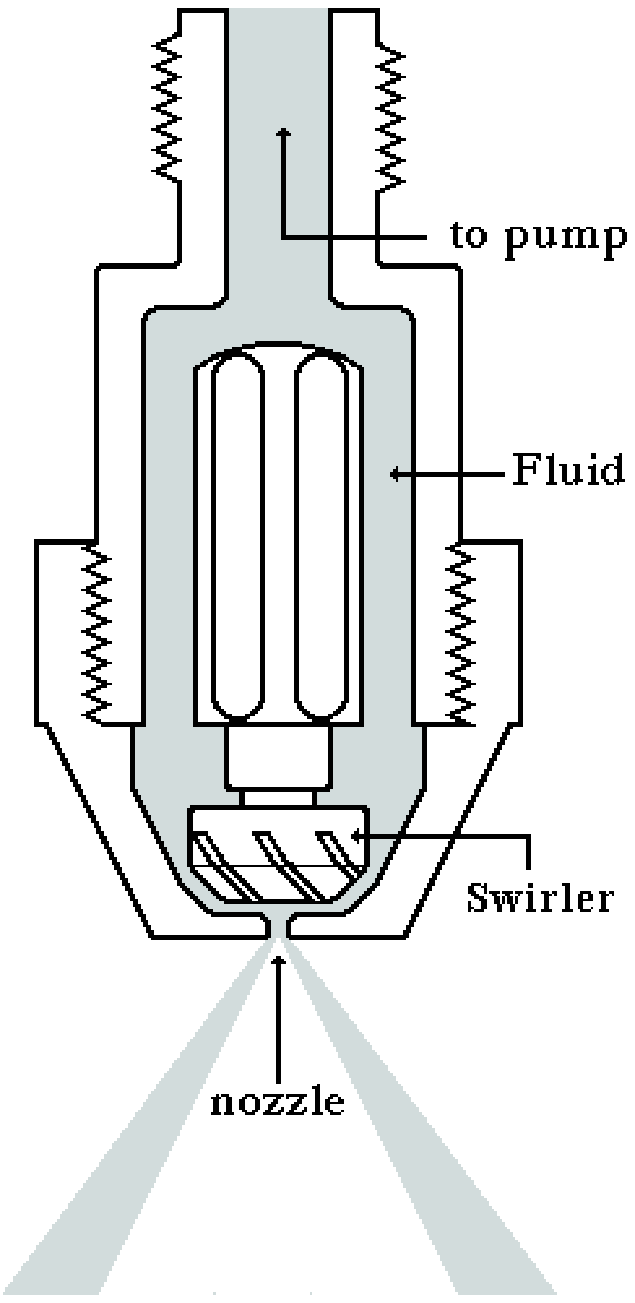}
\par
Fig. 2 ~Nozzle atomizer, a $1$ mm $\phi$ orifice producing a $45^{\circ}$ 
hollow cone.
\end{center}
\vskip-0.6cm
\end{figure}

\begin{figure}[ht]
\begin{center}
\epsfig{width=10cm,file=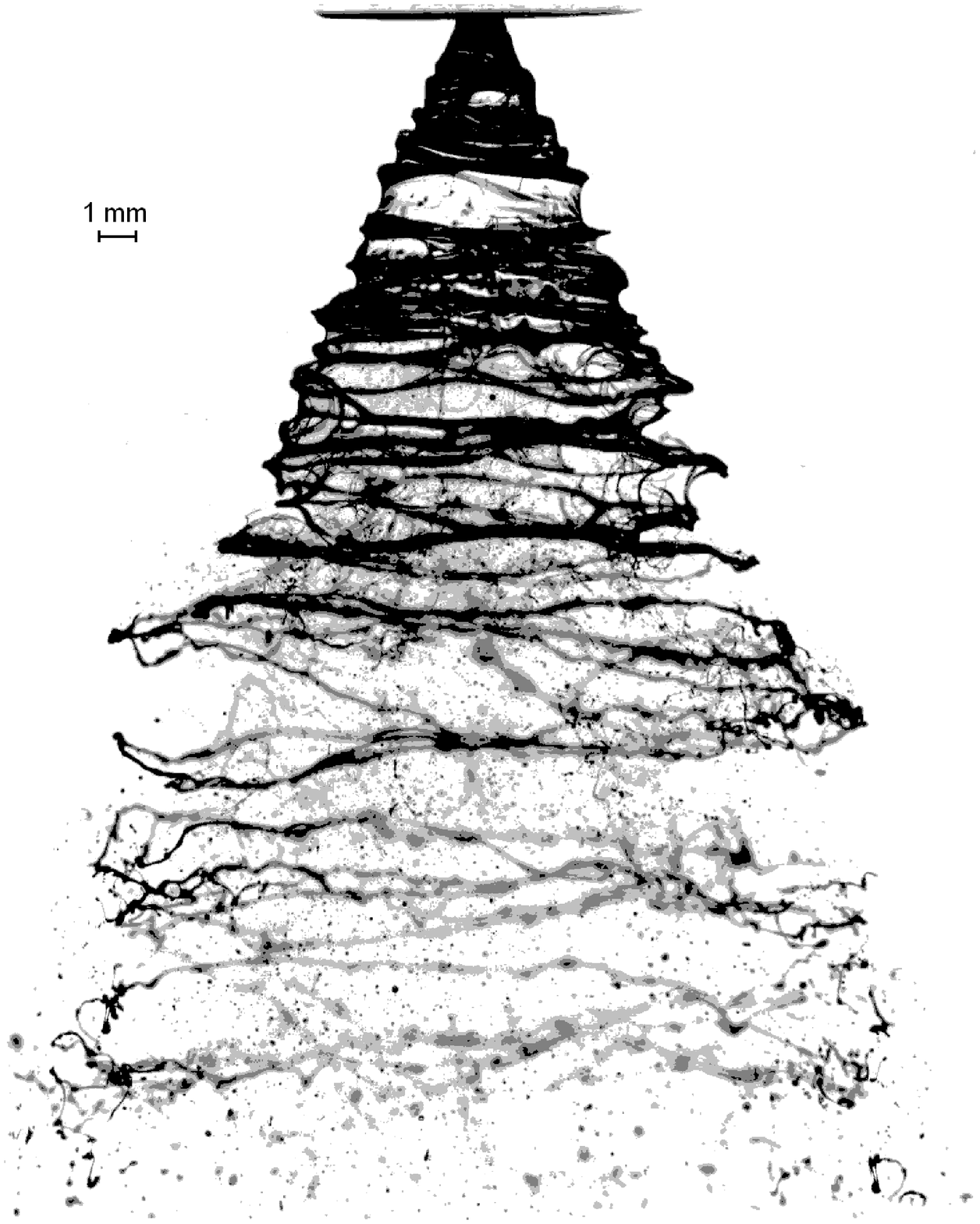}
\par
Fig. 3 ~Typical image obtained by the shadowgraph technique applied to an oil 
spray produced by the pressure swirl atomizer in Fig. 2.
\end{center}
\vskip-0.6cm
\end{figure}

\begin{figure}[ht]
\begin{center}
\epsfig{width=10cm,file=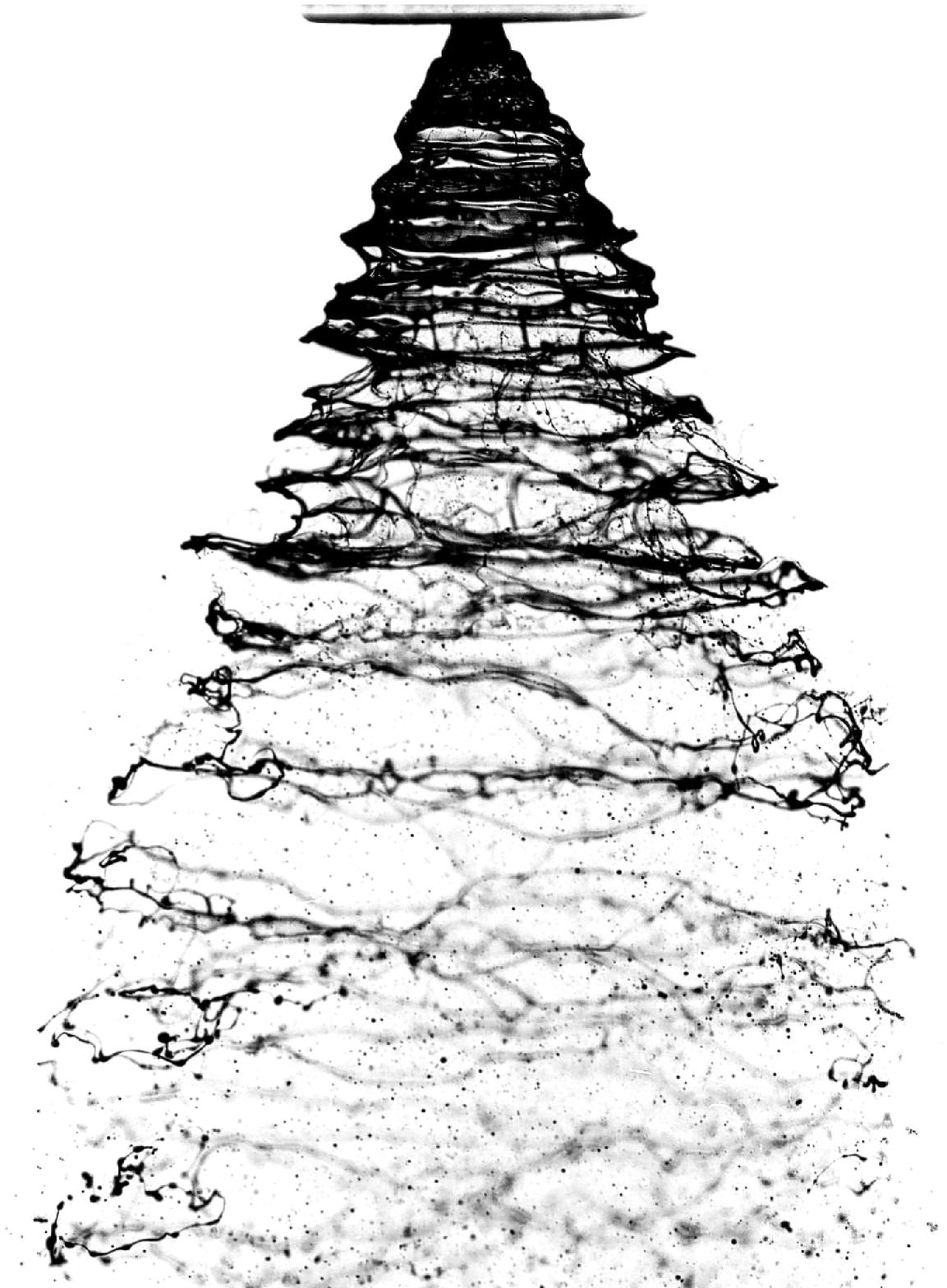}
\par
Fig. 4 ~Another typical image of the oil spray.
\end{center}
\vskip-0.6cm
\end{figure}

\begin{figure}[tbp]
\begin{center}
\epsfig{width=10cm,file=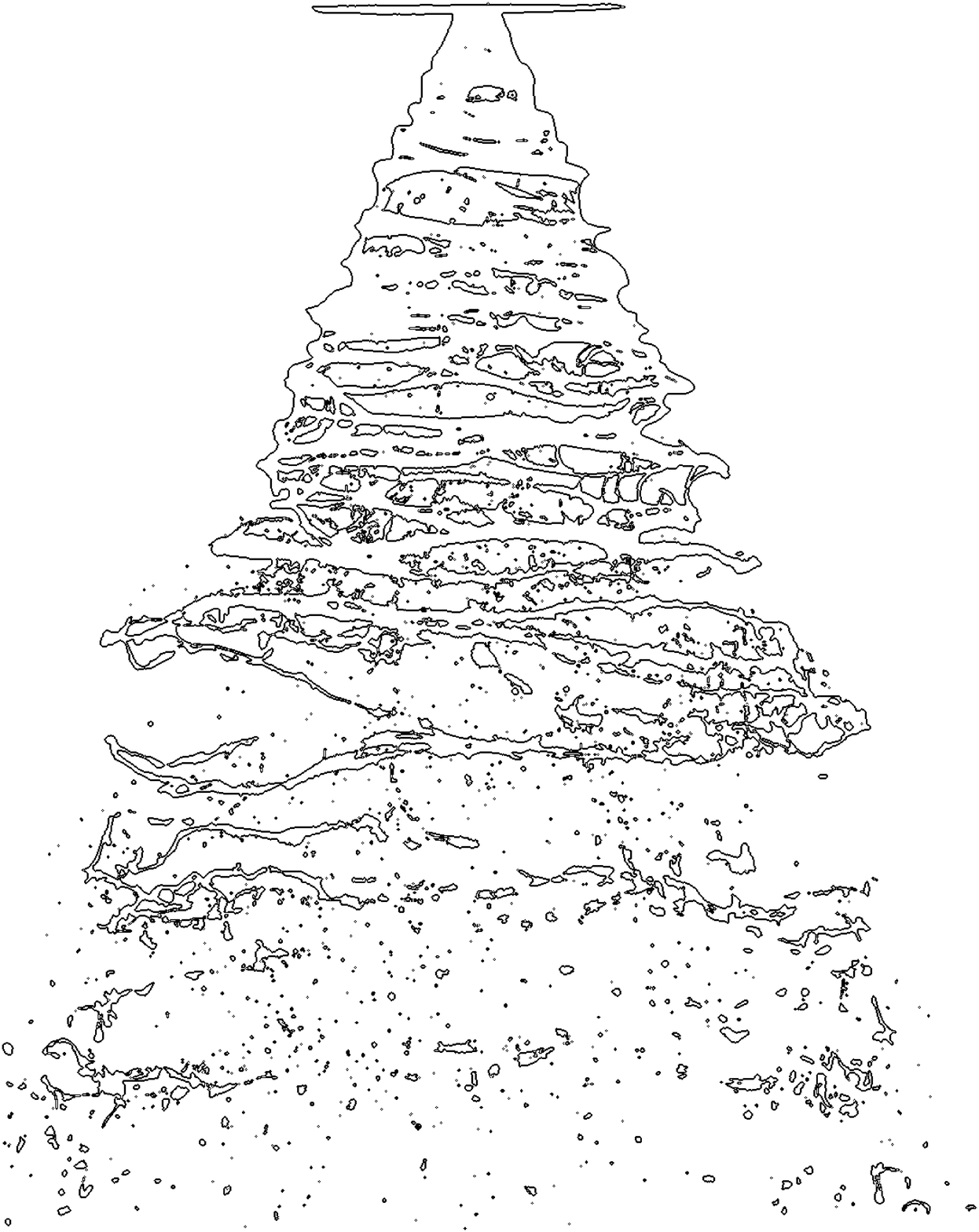}
\par
Fig. 5 ~Image showing the resulting contour of the spray in Fig. 3 when the 
filtering is done at half the maximum intensity value. 
\end{center}
\vskip-0.6cm
\end{figure}

\begin{figure}[tbp]
\begin{center}
\epsfig{width=15cm,file=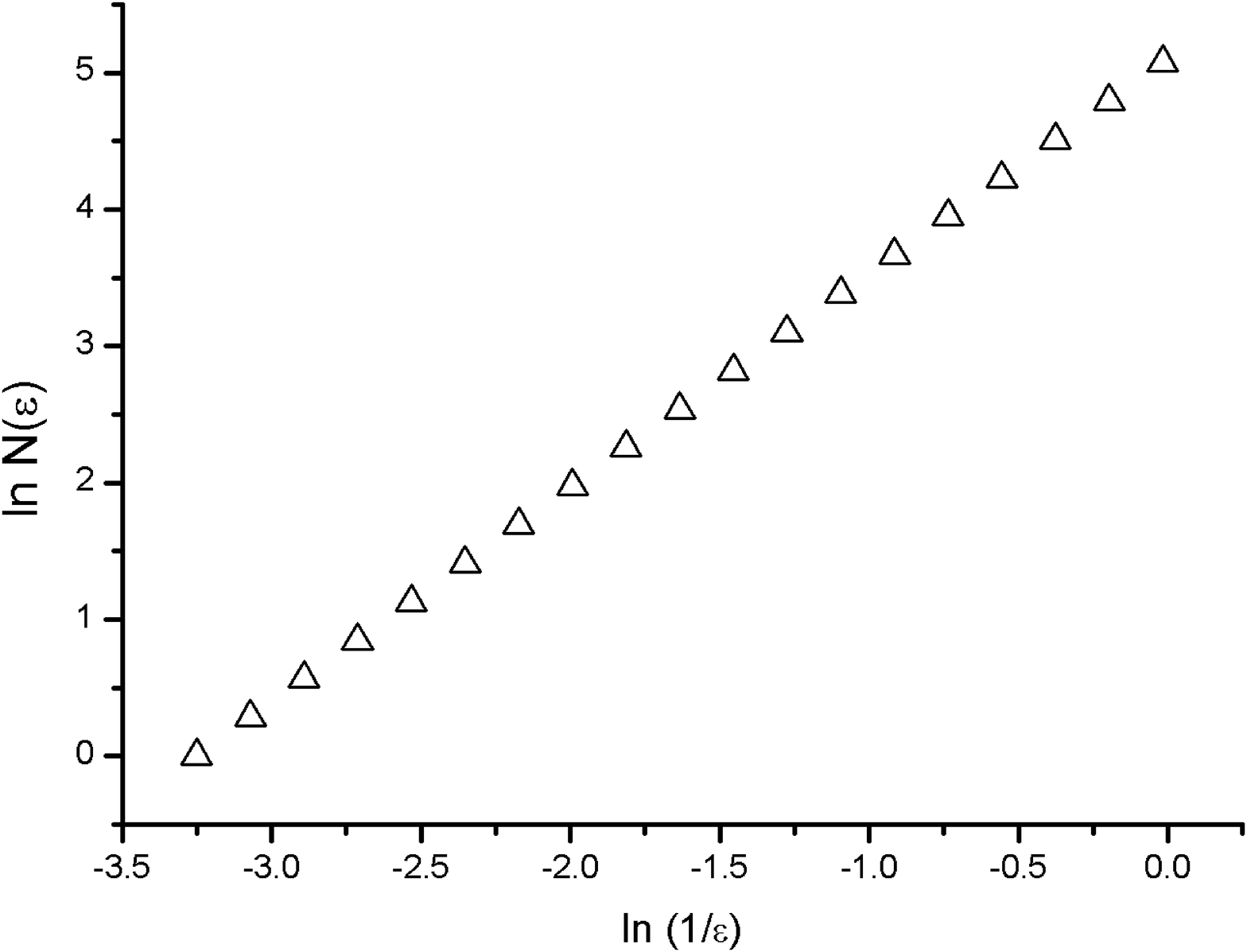}
\par
Fig. 6 ~Results from the Box-counting method at half-intensity filtering.
\end{center}
\vskip-0.6cm
\end{figure}

\begin{figure}[tbp]
\begin{center}
\epsfig{width=15cm,file=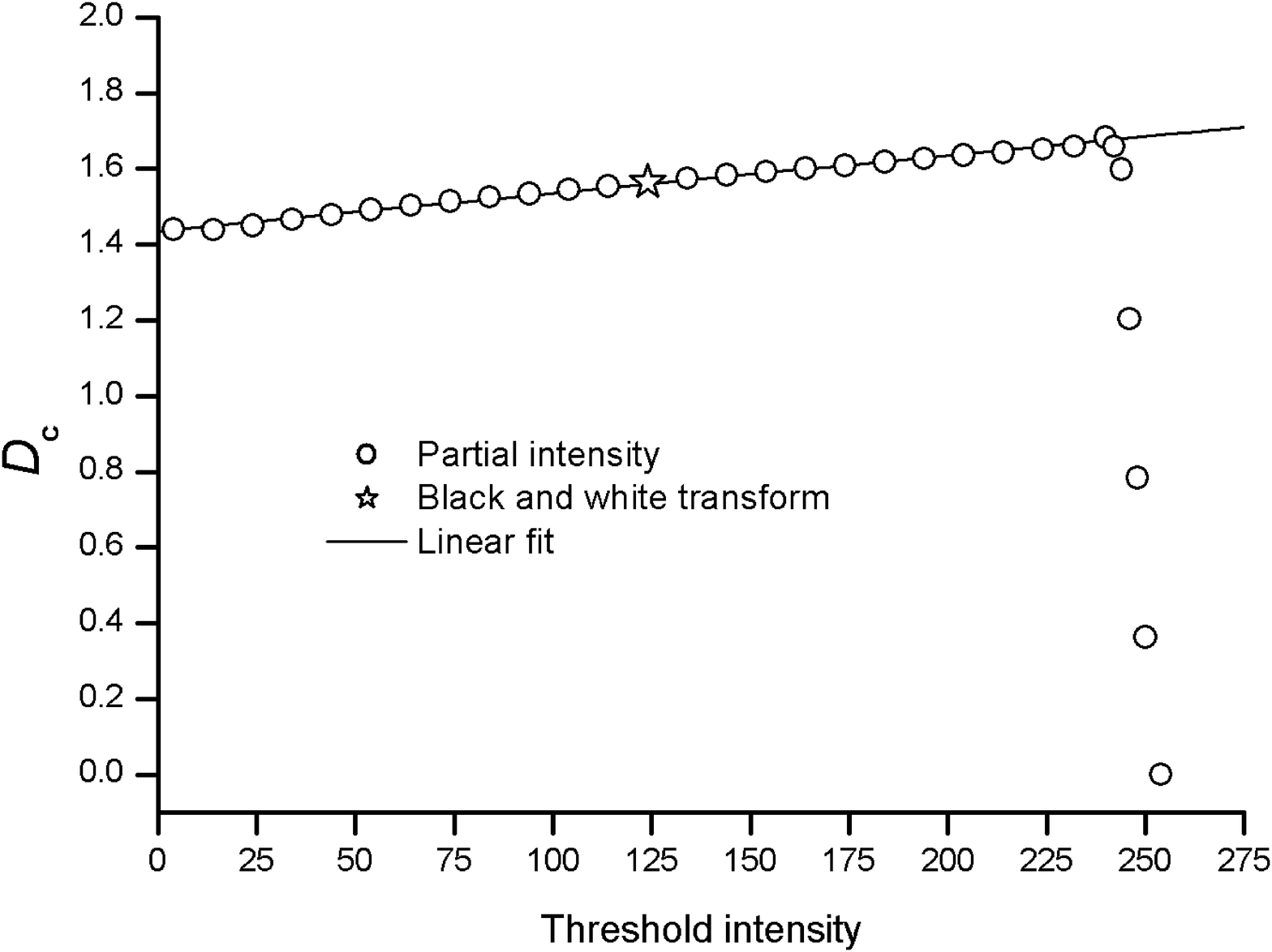}
\par
Fig. 7 ~Intensity threshold dependence of the fractal dimension.
\end{center}
\vskip-0.6cm
\end{figure}

\end{document}